\input epsf 
\documentstyle[preprint,eqsecnum,aps]{revtex}

\begin{document}

\count255=\time\divide\count255 by 60 \xdef\hourmin{\number\count255}
  \multiply\count255 by-60\advance\count255 by\time
 \xdef\hourmin{\hourmin:\ifnum\count255<10 0\fi\the\count255}

\preprint{\vbox{\hbox{WM-99-117}\hbox{JLAB-THY-99-31}
}}

\title{U(2) Flavor Physics without U(2) Symmetry}

\author{Alfredo Aranda$^a$\footnote{fefo@physics.wm.edu}, 
Christopher D.  Carone$^a$\footnote{carone@physics.wm.edu}, 
and Richard F. Lebed$^b$\footnote{lebed@jlab.org}}

\vskip 0.1in

\address{$^a$Nuclear and Particle Theory Group, Department of
Physics, College of William and Mary, Williamsburg, VA 23187-8795\\
$^b$Jefferson Lab, 12000 Jefferson Avenue, Newport News, VA 23606}
\vskip .1in
\date{October, 1999}
\vskip .1in

\maketitle
\tightenlines
\thispagestyle{empty}

\begin{abstract}
We present a model of fermion masses based on a minimal, non-Abelian 
discrete symmetry that reproduces the Yukawa matrices usually
associated with U(2) theories of flavor.  Mass and mixing 
angle relations that follow from the simple form of the quark and
charged lepton Yukawa textures are therefore common to both theories.
We show that the differing representation structure of our 
horizontal symmetry allows for new solutions to the solar and atmospheric
neutrino problems that do not involve modification of the original
charged fermion Yukawa textures, or the introduction of sterile neutrinos.
\end{abstract}


\newpage
\setcounter{page}{1}

\section{Introduction}\label{sec:intro}

One path toward understanding the observed hierarchy of fermion masses
and mixing angles is to assert that at some high energy scale all 
Yukawa couplings, except that of the top quark, are 
forbidden by a new symmetry $G_f$ that acts horizontally across 
the three standard model generations.  As this symmetry is spontaneously 
broken to smaller subgroups at successively lower energy scales, 
a hierarchy of Yukawa couplings can be generated.  The light fermion 
Yukawa couplings originate from  higher-dimension operators involving 
the standard model matter fields and  a set of `flavon' fields $\phi$, 
which are responsible for spontaneously breaking $G_f$.  The higher-dimension
operators are suppressed by a flavor scale $M_f$, which is the
ultraviolet cut-off of the effective theory; ratios of flavon vacuum 
expectation values (vevs) to the flavor scale, 
$\langle \phi \rangle/M_f$, provide a set of small symmetry-breaking
parameters that may be included systematically in the low-energy effective
theory.  Many models of this type have been proposed, with $G_f$ either 
gauged or global, continuous or discrete, Abelian or non-Abelian, or some
appropriate combination thereof~\cite{models}.  Non-Abelian symmetries 
are particularly interesting in the context of supersymmetric theories, where
flavor-changing neutral current (FCNC) processes mediated by
superparticle exchange can be phenomenologically unacceptable~\cite{mas}.
If the three generations of any given standard model matter field are
placed in {\bf 2}$\oplus${\bf 1} representations of some non-Abelian
horizontal symmetry group, it is possible to achieve an exact degeneracy 
between superparticles of the first two generations when $G_f$ is unbroken.  
In the low-energy theory, this degeneracy is lifted by the same small 
symmetry-breaking parameters that determine the light fermion Yukawa 
couplings, so that FCNC effects remain adequately suppressed, even with 
superparticle masses less than a TeV.

A particularly elegant model of this type  considered
in the literature assumes the continuous, global symmetry
$G_f=U(2)$~\cite{u21,u22,u23}.  Quarks and leptons are assigned 
to {\bf 2}$\oplus${\bf 1} representations, so that in tensor notation, 
one may represent the three generations of any matter field by $F^a+F^3$, 
where $a$ is a U(2) index, and $F$ is $Q$, $U$, $D$, $L$, or $E$.  
A set of flavons is introduced consisting of $\phi_a$, $S_{ab}$, and 
$A_{ab}$, where $\phi$ is a U(2) doublet, and  $S$ ($A$) is a 
symmetric (antisymmetric) U(2) triplet (singlet). If one assumes the 
pattern of vevs
\begin{equation}
\frac{\langle\phi\rangle}{M_f} = \left(\begin{array}{c} 0 
\\ \epsilon\end{array}\right),
\,\,\,\,\,
\frac{\langle S\rangle}{M_f} = \left(\begin{array}{cc} 0 & 0 
\\ 0 & \epsilon \end{array}
\right),
\,\,\,\,\,
\mbox{ and } 
\,\,\,\,\,
\frac{\langle A\rangle}{M_f}=
\left(\begin{array}{cc} 0 & \epsilon' \\ -\epsilon' & 0 \end{array}
\right) \,\,\, ,
\label{eq:vevs}
\end{equation}
which follows from the sequential breaking
\begin{equation}
U(2)\stackrel{\epsilon}{\rightarrow}U(1)\stackrel{\epsilon'}{\rightarrow}
 nothing \,\,\, ,
\end{equation}
then all fermion masses and Cabibbo-Kobayashi-Maskawa (CKM) mixing angles 
can be reproduced.  More specifically, the pattern of vevs 
in Eq.~(\ref{eq:vevs}) yields a Yukawa texture for the down quarks of 
the form
\begin{equation}
Y_D \sim \left(\begin{array}{ccc}  0 & d_1\epsilon' & 0 \\
-d_1\epsilon' & d_2\epsilon & d_3\epsilon \\
0 & d_4 \epsilon & 1 \end{array}\right) \,\,\, ,
\label{eq:yd}
\end{equation}
where $\epsilon\approx 0.02$, $\epsilon'\approx 0.004$, and 
$d_1,\ldots ,d_4$ are $O(1)$ coefficients that can be determined from 
Ref.~\cite{u23}.  Differences between hierarchies in $Y_D$ and
$Y_U$ can be obtained by embedding the model in a grand unified 
theory~\cite{u23}.  For example, in an SU(5) GUT, one obtains 
differing powers of $\epsilon$ and $\epsilon'$ in the up quark Yukawa 
matrix by assuming that $S_{ab}$ transforms as a {\bf 75}; combined GUT
and flavor symmetries prevent $A_{ab}$ and $S_{ab}$ from coupling
to the up and charm quark fields, unless an additional flavor singlet 
field $\Sigma$ is introduced that transforms as an SU(5) adjoint.  With 
$\langle \Sigma \rangle /M_f \sim \epsilon$, it is possible to
explain why $m_d :: m_s :: m_b = \lambda^4 :: \lambda^2 :: 1$, while 
$m_u :: m_c :: m_t = \lambda^8 :: \lambda^4 :: 1$, where $\lambda=0.22$ is
the Cabibbo angle.  The ratio $m_t/m_b$ is assumed to be unrelated to 
U(2) symmetry breaking, and is put into the low-energy theory by hand.  

In this letter we show that the properties of the U(2) model leading
to the successful Yukawa textures described above are also 
properties of smaller discrete symmetry groups.  To reproduce {\em all}
of the phenomenological successes of the U(2) model, we require 
a candidate discrete symmetry group to have the following properties: 
\begin{itemize}
\item {\bf 1}, {\bf 2}, and {\bf 3} dimensional representations. 
\item The multiplication 
rule {\bf 2}$\otimes${\bf 2}$=${\bf 3}$\oplus${\bf 1}. 
\item A subgroup $H_f$ such that the breaking pattern 
$G_f \rightarrow H_f \rightarrow nothing$ reproduces the
canonical U(2) texture given in Eq.~(\ref{eq:yd}).  This implies that
an unbroken $H_f$-symmetry forbids all Yukawa entries with $O(\epsilon')$
vevs, but not those with $O(\epsilon)$ vevs.
\end{itemize}
In the next section we show that the smallest group satisfying
these conditions is a  product of the double tetrahedral group $T'$ and an 
additional $Z_3$ factor.  Since U(2) is isomorphic to SU(2)$\times$U(1), 
it is not surprising that our candidate symmetry involves the product
of a discrete subgroup of SU(2), $T'$, and a discrete subroup of U(1), 
$Z_3$.  At this point, the reader who is unfamiliar with discrete group 
theory may feel somewhat uneasy.\footnote{For a review of basic terms,
see Ref.~\cite{d6}.} We stress that the group $T'$ is in fact 
a very simple discrete symmetry, a spinorial generalization of the 
symmetry of a regular tetrahedron (see Section II).  It is worth noting 
that the charge assignments in the model we present render $T'$ a 
nonanomalous discrete gauge symmetry, while the $Z_3$ factor is 
anomalous.  Models based on non-Abelian discrete gauge symmetries have 
yielded viable theories of fermion masses, as have models based on
discrete subgroups of anomalous U(1) gauge symmetries~\cite{models}.  In 
the latter case it is generally assumed that the U(1) anomalies are 
cancelled by the Green-Schwarz mechanism in string theory~\cite{GS}.  It 
is interesting that our model turns out to be a hybrid of these two ideas. 

One of the virtues of the model discussed in
this letter is that it allows for elegant extensions 
that explain the solar and atmospheric neutrino deficits, while
maintaining the original quark and charged lepton Yukawa textures.
This distinguishes our model from the modified version of the 
U(2) model presented in Ref.~\cite{caronehall}.  Preserving the 
U(2) charged fermion textures is desirable since they lead to successful 
mass and mixing angle relations such as $|V_{ub}/V_{cb}|=\sqrt{m_u/m_c}$, 
which are `exact' in the sense that they contain no unknown $O(1)$  
multiplicative factors.  Since we succeed in explaining solar
and atmospheric neutrino oscillations without sacrificing the
predictivity of the original model, we need not introduce sterile 
neutrinos, as in Ref.~\cite{hallweiner}.  However, we do not try to 
explain simultaneously the more controversial LSND results~\cite{lsnd}
in this paper.  We will consider versions of our model that include sterile 
neutrinos in a longer publication~\cite{nextpaper}.

\section{The Symmetry}

We seek a non-Abelian candidate group $G_f$ that provides the
{\bf 2}$\oplus${\bf 1} representation (rep) structure for the matter fields
described in the previous section. In order for the 
breaking of $G_f$ to reproduce the U(2) charged fermion Yukawa texture in
Eq.~(\ref{eq:yd}), one must have flavons that perform the same roles as
$\phi_a$, $S_{ab}$, and $A_{ab}$ in the U(2) model. Since these are
doublet, triplet, and nontrivial singlet reps, respectively, we require
$G_f$ to have reps of the same dimensions. Nontrivial singlets appear
in all discrete groups of order $<32$~\cite{TW}, so we seek groups $G_f$
with doublet and triplet representations.

        The order 12 tetrahedral group $T$, the group of proper
symmetries of a regular tetrahedron (which is also the alternating
group $A_4$, consisting of even permutations of four objects), is the
smallest containing a triplet rep, but has no doublet reps.  A number 
of groups with orders $< 24$ possess either doublet or triplet reps, but 
not both (See, for example, \cite{TW}).

        It turns out that two groups of order 24 possess both doublet
and triplet reps.  One is the symmetric group $S_4$ of permutations on
four objects, which is isomorphic to the group $O$ of proper
symmetries of a cube as well as the group $T_d$ of all proper and
improper symmetries of a regular tetrahedron.  $S_4$ possesses two 
triplets ${\bf 3}^\pm$, two singlets ${\bf 1}^\pm$, and one 
doublet ${\bf 2}$. However, in this case one encounters another difficulty:
The combination rule for doublets in $S_4$ is ${\bf 2} \otimes {\bf 2} 
= {\bf 2} \oplus {\bf 1}^- \oplus {\bf 1}^+$, which implies that
the triplet flavon cannot connect two doublet fields such as those of the 
first two generations of $Q$ and $U$. Thus, $S_4$ is not suitable for
our purposes.

        The unique group of order $< 32$ with the combination rule
${\bf 2} \otimes {\bf 2} \supset {\bf 3}$ is the double
tetrahedral group $T'$, which is order 24.  The character table, from
which one may readily generate explicit representation matrices, is
presented in Table~\ref{char}.  Geometrically, $T'$ is the group of
symmetries of a regular tetrahedron under proper rotations (Fig.~1).  
These symmetries consist of $1)$ rotations by $2\pi/3$ about an axis
connecting a vertex and the opposite face ($C_3$), $2)$ rotations by
$\pi$ about an axis connecting the midpoints of two non-intersecting
edges ($C_2$), and $3)$ the rotation $R$ by $2\pi$ about any axis,
which produces a factor $-1$ in the even-dimensional reps, exactly as
in SU(2).  Indeed, this feature is a consequence of $T^\prime \subset$
SU(2), and the rotations $C_3$ and $C_2$ are actually of orders 6 and
4, respectively.  Also, $T^\prime$ is isomorphic to the group
SL$_2$(F$_3$), which consists of 2 $\times$ 2 unimodular matrices
whose elements are added and multiplied as integers modulo 3.

        $T^\prime$ has three singlets ${\bf 1}^0$ and ${\bf 1}^\pm$,
three doublets, ${\bf 2}^0$ and ${\bf 2}^\pm$, and one triplet, {\bf
3}.  The {\it triality\/} superscript describes in a concise way the
rules for combining these reps: With the identification of $\pm$ as
$\pm 1$, the trialities add under addition modulo 3.  In addition, the
following rules hold:
\begin{equation}
\begin{array}{lcl}
{\bf 1} \otimes {\bf R} = {\bf R} \otimes {\bf 1}  =  {\bf R} 
\hspace{1em}\mbox{\rm for any rep ${\bf R}$},  &\hspace{3em} &
{\bf 2} \otimes {\bf 2}  =  {\bf 3} \oplus {\bf 1}, \\
{\bf 2} \otimes {\bf 3} = {\bf 3} \otimes {\bf 2}  =  {\bf 2}^0
\oplus {\bf 2}^+ \oplus {\bf 2}^- , &\hspace{3em} &
{\bf 3} \otimes {\bf 3}  =  {\bf 3} \oplus {\bf 3} \oplus {\bf 1}^0
\oplus {\bf 1}^+ \oplus {\bf 1}^- . \end{array} 
\end{equation}
Trialities flip sign under Hermitian conjugation.  Thus, for example,
${\bf 2}^+ \otimes {\bf 2}^- = {\bf 3} \oplus {\bf 1}^0$, and $({\bf
2}^+)^\dagger \otimes {\bf 2}^- = {\bf 3} \oplus {\bf 1}^+$.

One must now determine whether it is possible to place a
sequence of vevs hierarchically in the desired elements of the Yukawa
matrices.  Notice if $G_f$ is broken to a subgroup $H_f$ that rotates
the first generation matter fields by a common nontrivial phase, then $H_f$
symmetry forbids all entries with $O(\epsilon')$ vevs in 
Eq.~(\ref{eq:yd}). Therefore, we require that the elements of $G_f$
defining this subgroup have two-dimensional rep matrices
of the form {\it diag}$\{\rho,1\}$, with $\rho = \exp(2\pi i n/N)$ for some
$N$ that divides the order of $G_f$ and some integer $n$ relatively
prime with respect to $N$.  This form for $\rho$ follows because reps of 
finite groups may be chosen unitary, and must give the identity when raised
to the power of the order of $G_f$.  Such elements generate a subgroup
$H_f = Z_N$ of $G_f$.  Whether such elements exist in $G_f$ can
be determined since the rep of any element can be brought to diagonal form 
by a basis transformation, while the eigenvalues $\rho$, $1$ are invariant 
under such basis changes.

        Even if a given element $C \in G_f$ has the diagonal 
form {\it diag}$\{\rho_1, \rho_2\}$, $\rho_i = \exp( 2\pi i
n_i/N)$ (and thus generates a subgroup, $Z^C_N$, of $G_f$), a
phase rotation of the form {\it diag}$\{\rho,1\}$ can be achieved
if the original $G_f$ is extended by forming a direct product with 
an additional factor $Z_N$.  We then identify $H_f$ as a subgroup
of $Z^C_N \times Z_N$.  We choose one element of the additional $Z_N$ 
to compensate the phase of the 22 element of $C$, and similarly for 
the other elements of the $Z^C_N$. The element corresponding to $C$ 
in $G_f \times Z_N$ then effectively acts upon the doublet 
as {\it diag}$\{\exp[2\pi i (n_1 - n_2)/N],1\}$, and the remaining 
symmetry is $Z_{N/{\rm gcf}(N, |n_1 - n_2|)}$.  In the case 
that $|n_1 - n_2|$ and $N$ are relatively prime, this reduction amounts 
to forming the diagonal subgroup $Z^D_N$ of $Z^C_N \times Z_N$.  Similar 
arguments apply to the singlet and triplet reps.

In the particular case of $G_f=T'$, one finds elements $C$ that
generate either $Z_2$ or $Z_3$ subgroups.  By introducing an additional
$Z_n$ (with $n=2$ or $3$) one can arrange for a $Z_n$ subgroup that
affects only the first generation fields.  In the case of $Z_2$,
the nontrivial element of the diagonal subgroup is of the 
form {\it diag}$\{-1,1\}$, which leaves the 11 and 22 entries of the 
Yukawa matrices invariant. The incorrect relation $m_u=m_c$ then follows.  
On the other hand, $Z_3$ prevents an invariant 11 entry, so we are 
led to adopt
\begin{equation}
G_f = T^\prime \times Z_3 .
\end{equation}
        The reps of $G_f$ are named by extending the notation for
$T^\prime$ to include a superscript indicating the $Z_3$ rep.  These
are the trivial rep 0, which takes all elements to the identity, and
two complex-conjugate reps $+$ and $-$.  Like the trialities, these
indices combine via addition modulo 3.  We adopt the convention that
the $T^\prime \times Z_3$ reps ${\bf 1}^{00}$, ${\bf 1}^{+-}$,
${\bf 1}^{-+}$, ${\bf 2}^{0-}$, ${\bf 2}^{++}$ and ${\bf 2}^{-0}$
are special, in that these singlet reps and the second
component of the doublets remain invariant under $Z_3^D$.  Thus
any {\bf 2}$\oplus${\bf 1} combination of these reps is potentially
interesting for model building.

\section{A Minimal Model} \label{sec:model}

The minimal model has the three generations of matter fields transforming 
as ${\bf 2}^{0-} \oplus {\bf 1}^{00}$ under $G_f = T' \times Z_3$. The Higgs 
fields $H_{U,D}$ are pure singlets of $G_f$ and transform as 
${\bf 1}^{00}$. Given these assignments, it is easy to obtain the 
transformation properties of the Yukawa matrices,
\begin{eqnarray} \label{Yuk}
Y_{U,D,L} & \sim & \left( \begin{array}{c|c} 
[ {\bf 3}^{-}\oplus {\bf 1}^{0-}] 
& [ {\bf 2}^{0+}] \\ 
\hline
{[} {\bf 2}^{0+}]
 & [ {\bf 1}^{00} ]
\end{array} \right) .
\label{eq:ytp}
\end{eqnarray}
Eq.~(\ref{eq:ytp}) indicates the reps of the flavon fields 
needed to construct the fermion mass matrices. They are ${\bf 1}^{0-}$,
${\bf 2}^{0+}$, and  ${\bf 3}^{-}$, which we call $A$, $\phi$, and
$S$, respectively.  Once these flavons acquire vevs, the flavor group 
is broken. We are interested in a two-step breaking controlled by two 
small parameters $\epsilon$, and $\epsilon^{'}$, where
\begin{eqnarray} \label{break}
T' \otimes Z_3  \stackrel{\epsilon}{\longrightarrow} Z^D_{3} 
\stackrel{\epsilon^{'}}
{\longrightarrow} nothing \,\,\, .
\end{eqnarray}
Since we have chosen a doublet rep for the first two generations that
transforms as {\it diag}$\{\rho,1\}$ under $Z_3^D$, only the 22, 23, and
32 entries of the Yukawa matrices may develop vevs of $O(\epsilon)$, which
we assume originate from vevs in $S$ and $\phi$.  The symmetry $Z_3^D$
is then broken by a ${\bf 1}^{0-}$ vev of $O(\epsilon')$.  The 
Clebsch-Gordan coefficients that couple a ${\bf 1}^{0-}$ to two
${\bf 2}^{0-}$ doublets is proportional to $\sigma_2$, so the $\epsilon'$
appears in an antisymmetric matrix.  We therefore produce the U(2)
texture of Eq.~(\ref{eq:yd}).  Since the ${\bf 1}^{0-}$ and ${\bf 3}^-$
flavon vevs appear as antisymmetric and symmetric matrices,
respectively, all features of the grand unified extension of the U(2)
model apply here, assuming the same GUT transformation properties
are assigned to $\phi$, $S$, and $A$.  One can also show readily that 
the squark and slepton mass squared matrices are the same as in the U(2) 
model.

It is worth noting that we could construct completely equivalent theories
had we chosen to place the matter fields in reps like ${\bf 2}^{++}\oplus
{\bf 1}^{00}$ or  ${\bf 2}^{-0}\oplus{\bf 1}^{00}$, which have the
same transformation properties under $Z_3^D$ as our original choice.  The
reps ${\bf 2}^{0-}\oplus{\bf 1}^{00}$ are desirable in that they fill
the complete SU(2) representations ${\bf 2}\oplus {\bf 1}$, if we were to
embed $T'$ in SU(2).  Since anomaly diagrams linear in this SU(2) vanish
(and hence the linear Ib\'a$\tilde{{\rm n}}$ez-Ross condition is 
satisfied~\cite{IR}), we conclude that $T'$ is a consistent discrete 
gauge symmetry~\cite{BD}.  The additional $Z_3$ may also be considered 
a discrete gauge symmetry, providing its anomalies are cancelled by the 
Green-Schwarz mechanism.
\section{Neutrinos} \label{nus}

In this section, we show that the model presented in Section~\ref{sec:model}
can be extended to describe the observed deficit in solar and atmospheric
neutrinos. We consider two cases:

Case I: Here we do not assume grand unification, so that all flavons
are SU(5) singlets.  This case is of interest, for example, if one
is only concerned with explaining flavor physics of the lepton sector.
We choose
\begin{eqnarray} \label{right}
\nu_R \sim {\bf 2}^{0-} \oplus {\bf 1}^{-+} .
\end{eqnarray}
Note that the only difference from the other matter fields is the
representation choice for the third generation field. The neutrino Dirac 
and Majorana mass matrices then have different textures from the 
charged fermion mass matrices. Their transformation properties are
given by
\begin{equation}
M_{LR} \sim  \left( \begin{array}{c|c} [ {\bf 3}^{-} \oplus {\bf 1}^{0-}] 
& [ {\bf 2}^{+0}] \\ 
\hline
{[} {\bf 2}^{0+}]
 & [ {\bf 1}^{+-} ]
\end{array} \right) \,\,\, , \hspace{2em} 
M_{RR} \sim \left( \begin{array}{c|c} [ {\bf 3}^{-} ] 
& [ {\bf 2}^{+0}] \\ 
\hline
{[} {\bf 2}^{+0}]
 & [ {\bf 1}^{-+} ]
\end{array} \right) \,\,\, .
\end{equation}
Note that we obtain the same triplet and nontrivial singlet in the upper 
$2 \times 2$ block as in the charged fermion mass matrices, as well as
one of the same flavon doublets, the ${\bf 2}^{0+}$; the rep
${\bf 1}^{0-}$ is not present in $M_{RR}$, since Majorana mass matrices
are symmetric. In addition we obtain the reps 
${\bf 2}^{+0}$, ${\bf 1}^{+-}$, and ${\bf 1}^{-+}$, which did not appear 
in Eq.~(\ref{Yuk}). New flavon fields can now be introduced with these 
transformation properties, and their effects on the neutrino physics can 
be explored. Let us consider introducing a single\footnote{Assuming
more than one $\phi_\nu$ leads to the same qualitative results.} new 
flavon $ \phi_{\nu}$ transforming as a ${\bf 2^{+0}}$ and with 
a vev
\begin{eqnarray} \label{phinu}
\langle {\bf \phi_{\nu}} \rangle \sim \sigma_2
\left( \begin{array}{c} \epsilon^{'} \\
\epsilon \end{array} \right) \,\,\, ,
\label{eq:newvev}
\end{eqnarray}
where $\sigma_2$ is the Clebsch that couples the two doublets to 
${\bf 1}^{0-}$. The introduction of this new flavon is the only extension 
we make to the model in order to describe the neutrino phenomenology. After 
introducing $\phi_\nu$, the neutrino Dirac and Majorana mass matrices read
\begin{equation}
M_{LR}  =  \left( \begin{array}{ccc} 0 
& l_1 \epsilon^{'} & l_3 r_2 \epsilon^{'} \\ -l_1
\epsilon^{'} & l_2 \epsilon & l_3 r_1 \epsilon \\ 0 &
l_4 \epsilon & 0 \end{array} \right) \langle H_U \rangle \,\,\, ,
\hspace{2em}
M_{RR}  =  \left( \begin{array}{ccc} r_4 r_{2}^{2} \epsilon^{'2} 
& r_4 r_1 r_2 \epsilon \epsilon^{'} & r_2 \epsilon^{'} \\ r_4 r_1 r_2
\epsilon \epsilon^{'} & r_3 \epsilon & r_1 \epsilon \\ r_2 \epsilon^{'} &
r_1 \epsilon & 0 \end{array} \right) \Lambda_R \,\,\, ,
\end{equation}
where $\Lambda_R$ is the right-handed neutrino mass scale, and we have
parameterized the $O(1)$ coefficients.  Furthermore, the charged lepton
Yukawa matrix including $O(1)$ coefficients reads
\begin{eqnarray} \label{YL}
Y_{L}  \sim  \left( \begin{array}{ccc} 0 & c_1 \epsilon^{'} & 0 
 \\ -c_1 \epsilon^{'} & 3c_2\epsilon & c_3\epsilon 

 \\ 0 & c_4\epsilon & 1
\end{array} \right) \,\,\, .
\end{eqnarray}
The factor of $3$ in the 22 entry is simply assumed at present, but 
originates from the Georgi-Jarlskog mechanism~\cite{GJ} in the
grand unified case considered later.

The left-handed Majorana mass matrix $M_{LL}$ follows from the 
seesaw mechanism  
\begin{eqnarray} \label{seesaw}
M_{LL} \approx M_{LR} M_{RR}^{-1} M_{LR}^{T} \,\,\, ,
\end{eqnarray}
which yields
\begin{eqnarray} \label{MLL}
M_{LL}  \sim  \left( \begin{array}{ccc} (\epsilon'/\epsilon)^2 & 
\epsilon'/\epsilon & \epsilon'/\epsilon 
\\ \epsilon'/\epsilon & 1 & 1 
\\ \epsilon'/\epsilon & 1 & 1
\end{array} \right) \frac{\langle H_U \rangle^2 \epsilon}{\Lambda_R} ,
\end{eqnarray}
where we have suppressed the $O(1)$ coefficients. We naturally obtain 
large mixing between second- and third-generation
neutrinos, while the 12 and 13 mixing angles are $O(\epsilon'/\epsilon)$.
However, taking into account the diagonalization of $Y_L$, the relative
12 mixing angle can be made smaller, as we discuss below.  Explanation of
the observed atmospheric neutrino fluxes by $\nu_\mu$-$\nu_\tau$ mixing
suggests $\sin^2 2\theta_{23} \agt 0.8$ and $10^{-3}\alt \Delta m^2_{23} \alt 
10^{-2}$, while the solar neutrino deficit may be accommodated assuming the 
small-angle MSW solution $2\times 10^{-3} \alt \sin^2 2\theta_{12} 
\alt 10^{-2}$ for $4\times10^{-6} \alt \Delta m^2_{12} \alt 10^{-5}$, where 
all squared masses are given in eV$^2$.  We display below an explicit choice 
of the $O(1)$ parameters that yields both solutions simultaneously; a more 
systematic global fit will be presented in Ref.~\cite{nextpaper}.

If $M_{LL}$ and $Y_{L}$ are diagonalized by 
$M_{LL} = V M_{LL}^{0} V^{\dagger}$,
$Y_{L} = U_{L} Y_{L}^{0} U_{R}^{\dagger}$, then the neutrino CKM matrix is
given by
\begin{eqnarray} \label{CKM}
V_{\nu} = U_{L}^{\dagger} V .
\end{eqnarray}
We aim to reproduce the 12 and 23 mixing angles, as well as the ratio 
$10^2 \alt \Delta m^2_{23}/\Delta m^2_{12} \alt 2.5 \times 10^3$
suggested by the data.  Obtaining this ratio is sufficient since $\Lambda_R$
is not determined by symmetry considerations and may be chosen freely.
Assuming the previous values  $\epsilon = 0.02$ and 
$\epsilon^{'} = 0.004$ and the parameter set 
$(l_1,\ldots, l_4, r_1,\ldots, r_4, c_1,\ldots, c_4)=$ 
$(0.5, 1.0, -1.2, 2.3, 1.0, 1.0, 1.0, 1.0, 1.0, 1.0, 1.0, 1.0)$, we find:
\begin{equation}
\frac{\Delta m_{23}^{2}}{\Delta m_{12}^{2}} = 105 ,
\,\,\,\,\,\,\,\,\,\,
{\sin^{2} 2 \theta_{12}} = 5 \times 10^{-3} ,
\,\,\,\,\,\,\,\,\,\,
{\sin^{2} 2 \theta_{23}} = 0.9 ,  
\end{equation}
which fall in the desired ranges.  While all our coefficients are of
natural size, we have arranged for an $O(15\%)$ cancellation between
12 mixing angles in $U_L$ and $V$ to reduce the size of
$\sin^2 2\theta_{12}$ to the desired value.

Case II: Here we assume that the flavons transform nontrivially under
an SU(5) GUT group, namely $A \sim {\bf 1}$, $S \sim {\bf 75}$,
$\phi \sim {\bf 1}$, and $\Sigma \sim{\bf 24}$. Note that since 
$\overline{H} \sim {\bf \overline{5}}$, the products $S\overline{H}$ 
and $A \overline{H}$ transform as a ${\bf \overline{45}}$ 
and ${\bf \overline{5}}$, respectively, ultimately providing a factor 
of $3$ enhancement in the 22 entry of $Y_{L}$ (the Georgi-Jarlskog 
mechanism). In addition, two ${\bf 2}^{+0}$ doublets are 
introduced, $\phi_{\nu 1}$ and $\phi_{\nu 2}$, since the texture 
obtained for the neutrino masses by adding only one 
extra doublet is not viable.  Both doublets $\phi_\nu$ have vevs of 
the form displayed in Eq.~(\ref{eq:newvev}).  Crucially, the presence of
these two new doublets does not alter the form of any charged fermion
Yukawa texture.

The neutrino Dirac and Majorana mass matrices now take the form
\begin{equation}
M_{LR}  = \left( \begin{array}{ccc} 0 
& l_1 \epsilon^{'} & l_5 r_2 \epsilon^{'} \\ -l_1
\epsilon^{'} & l_2 \epsilon^{2} & l_3 r_1 \epsilon \\ 0 &
l_4 \epsilon & 0 \end{array} \right) \langle H_U \rangle \,\,\, ,
\hspace{2em}
M_{RR}  =  \left( \begin{array}{ccc} r_3 \epsilon^{'2} 
& r_4 \epsilon \epsilon^{'} & r_2 \epsilon^{'} \\ r_4
\epsilon \epsilon^{'} & r_5 \epsilon^{2} & r_1 \epsilon \\ r_2 \epsilon^{'} &
r_1 \epsilon & 0 \end{array} \right) \Lambda_R \,\,\, ,
\end{equation}
while the charged fermion mass matrix is the same as in Eq.~(\ref{YL}). 
Using Eq.~(\ref{seesaw}) one obtains the texture:
\begin{eqnarray} \label{MLLGUT}
M_{LL}  \sim  \left( \begin{array}{ccc} (\epsilon'/\epsilon)^{2} & 
\epsilon{'}/\epsilon & \epsilon{'}/\epsilon 
\\ \epsilon{'}/\epsilon & 1 & 1 
\\ \epsilon{'}/\epsilon & 1 & 1
\end{array} \right) \frac{\langle H_U \rangle^2}{\Lambda_R} \,\,\, .
\end{eqnarray}
If we now choose $(l_1,\ldots, l_5, r_1,\ldots, r_5, c_1,\ldots, c_4)=$ 
$(-1.0,$ $1.0,$ $1.0,$ $0.5,$ $1.0,$ $0.5,$ $1.0,$ $1.0,$ $1.0,$ $-2.0,$ 
$1.0,$ $1.0,$ $1.0,$ $1.0)$,
we find
\begin{equation}
\frac{\Delta m_{23}^{2}}{\Delta m_{12}^{2}} = 282 ,
\,\,\,\,\,\,\,\,\,\,
{\sin^{2} 2 \theta_{12}} = 6 \times 10^{-3} ,
\,\,\,\,\,\,\,\,\,\,
{\sin^{2} 2 \theta_{23}} = 0.995 .  
\end{equation}
Again these values fall in the desired ranges to explain the
atmospheric and solar neutrino deficits, assuming an appropriate
choice for $\Lambda_R$.

\section{Conclusions}

In this letter we have shown how to reproduce the quark and 
charged lepton Yukawa textures of the U(2) model in their entirety,
using a minimal non-Abelian discrete symmetry group.  We showed that 
the representation structure of $T'\times Z_3$, in particular the 
existence of three distinct  2-dimensional irreducible representations, 
allows for solutions to the solar and atmospheric neutrino problems that 
require neither a modification of the simple charged fermion Yukawa textures 
of the U(2) model nor the introduction of singlet neutrinos.  The 
simplicity of the symmetry structure of our model suggests that a more 
comprehensive investigation of the space of possible models is justified.
Work on alternative neutrino sectors as well as a more detailed 
phenomenological analysis of the models described here will be presented 
elsewhere~\cite{nextpaper}.

{\samepage
\begin{center}
{\bf Acknowledgments}
\end{center}
AA and CC thank the National Science Foundation for support under 
Grant Nos.\ PHY-9800741 and PHY-9900657.  RFL thanks the Department of 
Energy for support under Contract No.\ DE-AC05-84ER40150.}

\begin{table}
\begin{tabular}{c|ccccccc}
Sample element & $E$ & $R$ & $C_2, C_2 R$ & $C_3$ & $C_3^2$ & $C_3 R$
& $C_3^2 R$ \\ \hline
Order of class & 1 & 1 & 6 & 4 & 4 & 4 & 4 \\

Order of element & 1 & 2 & 4 & 6 & 3 & 3 & 6 \\ \hline  

${\bf 1}^0$ & 1 & 1 & 1 & 1 & 1 & 1 & 1 \\

${\bf 1}^+$ & 1 & 1 & 1 & $\eta$ & $\eta^2$ & $\eta$ & $\eta^2$ \\

${\bf 1}^-$ & 1 & 1 & 1 & $\eta^2$ & $\eta$ & $\eta^2$ & $\eta$ \\

${\bf 2}^0$ & 2 & $-2$ & 0 & 1 & $-1$ & $-1$ & 1 \\

${\bf 2}^+$ & 2 & $-2$ & 0 & $\eta$ & $-\eta^2$ & $-\eta$ & $\eta^2$ \\

${\bf 2}^-$ & 2 & $-2$ & 0 & $\eta^2$ & $-\eta$ & $-\eta^2$ & $\eta$ \\

${\bf 3}$ & 3 & 3 & $-1$ & 0 & 0 & 0 & 0
\end{tabular}
\caption{Character table of the double tetrahedral group $T'$.  The
phase $\eta$ is $\exp(2\pi i/3)$.}
\label{char}
\end{table}

\begin{figure}
  \begin{centering}
  \def\epsfsize#1#2{1.0#2}
\hfil\hspace{-10em} \epsfbox{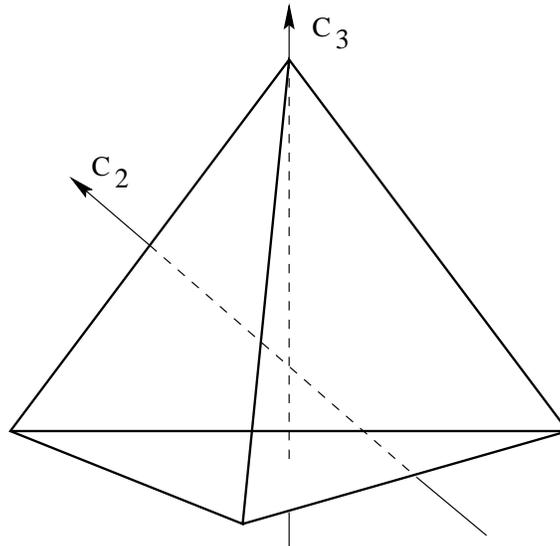} \hfill
\caption{Geometrical illustration of the group $T'$.  The rotations
$C_2$ and $C_3$ are defined in the text.}
  \end{centering}
\end{figure}

\end{document}